\documentclass{jpsj3}
\usepackage{txfonts,tabularx}
\usepackage[usenames]{color}

\title{Unscaling Superconducting Parameters with $T_c$ for Bi-2212 and Bi-2223:
\\A Magnetotransport Study in the Superconductive Fluctuation Regime}

\author{Shintaro ADACHI$^1$\thanks{E-mail address: bi2223hu@gmail.com},
 Tomohiro USUI$^1$,
 Yasuhito ITO$^1$,
 Hironobu KUDO$^1$,
 Haruki KUSHIBIKI$^1$,
 Kosuke MURATA$^1$,
 Takao WATANABE$^1$,
 Kazutaka KUDO$^2$\thanks{Present address: Department of Physics, Okayama University, Okayama 700-8530, Japan.},
 Terukazu NISHIZAKI$^2$\thanks{Present address: Department of Electrical Engineering and Information Technology, Kyushu Sangyo University, Fukuoka 813-8503, Japan.},
 Norio KOBAYASHI$^2$,
 Shojiro KIMURA$^2$,
 Masaki FUJITA$^2$,
 Kazuyoshi YAMADA$^2$\thanks{Present address: Institute of Materials Structure Science, KEK, Tsukuba 305-0801, Japan.},
 Takashi NOJI$^3$,
 Yoji KOIKE$^3$,
 and Takenori FUJII$^4$}

\inst{$^1$Graduate School of Science and Technology, Hirosaki University, Hirosaki 036-8561, Japan \\
$^2$Institute for Materials Research, Tohoku University, Sendai 980-8577, Japan \\
$^3$Graduate School of Engineering, Tohoku University, Sendai 980-8579, Japan \\
$^4$University of Tokyo, Cryogenic Research Center, Tokyo 113-0032, Japan} 

\abst{To investigate the origin of the enhanced $T_c$ ($\approx$ 110 K) of the trilayer cuprate superconductor Bi$_{2}$Sr$_{2}$Ca$_{2}$Cu$_{3}$O$_{10+\delta}$ (Bi-2223), we have performed systematic magnetoresistance (MR) measurements on this superconductor, as well as on the bilayer superconductor, Bi$_{2}$Sr$_{2}$CaCu$_{2}$O$_{8+\delta}$ (Bi-2212). The in-plane coherence length, $\xi_{ab}$, and the specific-heat jump, ${\Delta}C$, have been estimated using the theory of renormalized superconductive fluctuations, and the doping dependence of these parameters has been qualitatively explained using the Fermi arc approach. A detailed comparison of the superconducting parameters with $T_c$ for these compounds suggests that an additional superconducting condensation energy exists, due to an increase in the number of stacking CuO$_{2}$ planes in a unit cell.}

\begin{document}
\maketitle

\section{Introduction}

It is desired that the superconducting transition temperature, $T_c$, of high-$T_c$ cuprates be further increased in order to extend their field of potential applications. It is empirically known that $T_c$ increases with an increased number, $n$, of CuO$_{2}$ planes in a unit cell~\cite{karppinen}. Recently, nuclear magnetic resonance (NMR) measurements have revealed~\cite{mukuda1} that long range antiferromagnetic order and high-$T_c$ superconductivity coexists in a single CuO$_{2}$ plane for multilayered ($n \ge$ 3) high-$T_c$ cuprates, thereby implying that they are intimately related. This finding has greatly motivated the study of the microscopic mechanism behind this response. 

In order to examine this process in greater detail, the ``Bi family", i.e., Bi$_{2}$Sr$_{2}$CuO$_{6+\delta}$ (Bi-2201: $n$ = 1), Bi$_{2}$Sr$_{2}$CaCu$_{2}$O$_{8+\delta}$ (Bi-2212: $n$ = 2), and Bi$_{2}$Sr$_{2}$Ca$_{2}$Cu$_{3}$O$_{10+\delta}$ (Bi-2223: $n$ = 3), is the best choice, since large and high-quality Bi-2223 single crystals grown using the traveling solvent floating zone (TSFZ) method are now available~\cite{fujii}. To date, several spectroscopic measurements, including angle-resolved photoemission spectroscopy (ARPES)~\cite{sato,matsui,ideta,feng1} and interlayer tunneling spectroscopy (ITS)~\cite{suzuki3}, have been performed on Bi-2223 single crystals. All of these studies have reported a larger superconducting gap, $\Delta_{SG}$, for Bi-2223 than those of Bi-2212 or Bi-2201. Sato et al.~\cite{sato,matsui} have shown that $\Delta_{SG}$ scales with $T_c$ among the optimally doped Bi-family compounds, pointing out that the greater pairing strength is responsible for the higher $T_c$ of Bi-2223. On the other hand, Feng et al.~\cite{feng1} have observed that $\Delta_{SG}$ and the single-particle coherent weight, $z_{A}$, which is regarded as a measure of the superfluid density, $\rho_{s}$ (or phase stiffness)~\cite{feng}, both scale linearly with $T_c$. Based on this observation, they strongly emphasized the importance of the phase stiffness in raising $T_c$, which is consistent with the original idea of Emery and Kivelson~\cite{emery} and the well known ``Uemura's plot"~\cite{uemura,uemura1}. Therefore, it is very important to check this scaling using different procedures, and to investigate whether or not the scaling relations among different $n$-values can be extended to non-optimal doping states.

In the simple BCS theory, $\Delta_{SG}$ is correlated with the in-plane coherence length, $\xi_{ab}$, via $\xi_{ab}\propto\hbar{v_{F}}/\Delta_{SG}$, where $v_{F}$ is the in-plane Fermi velocity. Therefore, we can regard $\xi_{ab}^{-1}$ as being a measure of $\Delta_{SG}$. Conventionally, $\xi_{ab}$ is obtained by the direct measurement of the upper critical field, $H_{c2}$, at the lowest temperature, or by an extrapolation of $H_{c2}$ near $T_c$ to absolute zero using the Werthamer-Helfand-Hohenberg (WHH) theory~\cite{WHH}. However, the $H_{c2}$ of high-$T_c$ cuprates in many cases exceeds 100 T and, thus, we cannot determine this value directly. The resistive transition shows the broadening behavior under a magnetic field and, therefore, the WHH theory cannot be applied. In high-$T_c$ cuprates, $\xi_{ab}$ is extremely short (10 - 30 {\AA}) and $T_c$ is high. Furthermore, strong magnetic fields applied perpendicular to CuO$_{2}$ planes quantize the orbitals of the Cooper pairs to Landau levels, which causes a reduction in the fluctuation-freedom by two dimensions, meaning that, for extremely two-dimensional high-$T_c$ cuprates, the fluctuation-freedom effectively becomes zero-dimensional. All these facts strongly enhance the superconductive fluctuations, therefore the non-linear term ($|\psi|^{4}$ term, with $\psi$ the order parameter) in the Ginzburg-Landau (GL) free energy cannot be ignored, even for temperatures other than $T_c$. Ikeda, Ohmi, and Tsuneto (IOT) have succeeded in describing this ``critical fluctuation" region, taking the non-linear term into account, and have interpreted the experimentally observed broadening behavior in the resistive transitions very well~\cite{ikeda}. Because this theory contains $\xi_{ab}$ as a parameter, we can obtain the value of this term by numerically fitting the resistive transition curves under the magnetic fields~\cite{wataB,semba1}.
The superfluid density, $\rho_{s}$, is proportional to $1/\lambda_{L}^{2}$, where $\lambda_{L}$ is the in-plane London penetration depth, and $\lambda_{L}$ is correlated with the specific-heat jump, $\Delta{C}$, via the GL formula $\lambda_{L}/\xi_{ab} = \phi_{0}/(2\pi\xi_{ab}^{2}\sqrt{8\pi{T_{c0}}\Delta{C}})$~\cite{ikeda}. Here, $\phi_{0}$ is the flux quantum and $T_{c0}$ is the mean-field superconducting transition temperature. Therefore, $\rho_{s}$ can be estimated if we know $\Delta{C}$. Since the excess conductivity due to the superconductive fluctuation is reflected in the appearance of the order parameter amplitude, $\langle|\psi|^{2}\rangle$, due to the thermodynamic fluctuations, $\Delta{C}$, which is a typical thermodynamic quantity accompanied by the superconducting transition, naturally included in the IOT theory~\cite{ikeda}. Therefore, the value of $\Delta{C}$ is also obtained by fitting the resistive transition curves under the magnetic fields~\cite{wataB,semba1}.   

In this paper, we measure the in-plane resistive transition for variously doping-controlled trilayer Bi-2223, as well as bilayer Bi-2212, under various magnetic fields, $B$, parallel to the $c$-axis, ($\mbox{\boldmath$B$}\parallel{c}$). The data are analyzed using the superconducting-fluctuation-renormalized Ginzburg-Landau (GL) theory (IOT theory)~\cite{ikeda}. Subsequent analysis enables us to estimate superconducting parameters $\xi_{ab}$ and $\Delta{C}$. By using these parameters, $1/\lambda_{L}^{2}$ ($ \propto \rho_{s}$) is estimated. Then, $\xi_{ab}$ and $1/\lambda_{L}^{2}$ are compared with previously reported values and the doping dependence of these parameters is determined. Finally, we show the correlations of these parameters with $T_c$ for Bi-2212 and Bi-2223, and discuss the implications of these results. 

\section{Experimental Details}

High-quality single Bi-2212~\cite{usu} and Bi-2223 crystals were grown using the TSFZ method. The Bi-2212 samples used here are the same as those reported in ref. 17, apart from one (sample B), while the Bi-2223 crystals were grown under slightly different conditions to those of earlier reports~\cite{fujii,t.fujii}, details of which will be published elsewhere. The Bi-2223 crystals were then subjected to annealing in a similar manner to that given in ref. 18, in order to variously control their doping levels. However, the precise doping levels, $p$, were not determined, because Bi-2223 has two crystallographically inequivalent CuO$_{2}$ planes with different $p$ ~\cite{mukuda}, and thus an empirical relation~\cite{tallon} could not be applied. Instead, we simply judged the optimally doped level by monitoring the maximum $T_c$. Further doping caused a slight reduction in $T_c$, which is somewhat different from the findings of our earlier report~\cite{t.fujii}. This difference may be attributed to the adoption of different growth conditions. The $\rho_{ab}$ measurements were carried out using a DC four-terminal method, and magnetic fields, $B$, of up to 17.5 T were applied parallel to the $c$-axis with a superconducting magnet.

The numerical analysis was performed in two steps. First, the zero-field, $\rho_{ab}$, data was analyzed using the formula, $\rho_{ab} = 1/(\rho_{n}^{-1} + \sigma_{2D-AL})$, where $\rho_{n}$ is the in-plane resistivity when superconductive fluctuation effects are absent and $\sigma_{2D-AL}$ is the zero-field two-dimensional (2D) Aslamazov-Larkin (AL) form~\cite{AL} for the superconductive fluctuation. (Here, $\sigma_{2D-AL}$ = $e^{2}\epsilon^{-1}$/16$\hbar{d}$, where $\epsilon$ is the reduced temperature and $\epsilon = \ln(T/T_{c0})$, while $d$ is the interlayer spacing.) The $d$ values were set to 15.4 and 18.5 {\AA} for Bi-2212 and Bi-2223, respectively. In this study, $\rho_{n}$ is simply assumed to be $\rho_{n} = a{T} +b$. We did not use a $C$ factor (i.e., $C$ = 1)~\cite{oh}, which is defined by the ratio of the actual $\rho_{ab}$ of an imperfect crystal to an ideal crystal and, thus, it phenomenologically adjusts the magnitude of the fluctuation conductivity ($\sigma_{2D-AL} \to \sigma_{2D-AL}/C$). We optimized $T_{c0}$ and $\rho_{n}$ to reproduce the zero-field data.
Next, the in-plane resistive transitions under various magnetic fields were analyzed using the expression for the excess conductivity, $\sigma^{fl}$, derived from the IOT theory~\cite{ikeda}. The value of the out-of-plane coherence length, $\xi_{c}$, was set to 0.1 {\AA} for all samples, because the theoretical curve is insensitive in the region of this value~\cite{wataB}. In the analysis, we first optimized $\Delta{C}$ to reproduce the resistivity data roughly.  Then, we optimized $\xi_{ab}$ to reproduce the data well, and this process was repeated until a satisfactory fit was obtained. In this way, we systematically estimated $\Delta{C}$ and $\xi_{ab}$ for Bi-2212 and Bi-2223 with their doping levels variously controlled.

\section{Results and Discussion}

\begin{figure*}
\begin{center}
\includegraphics{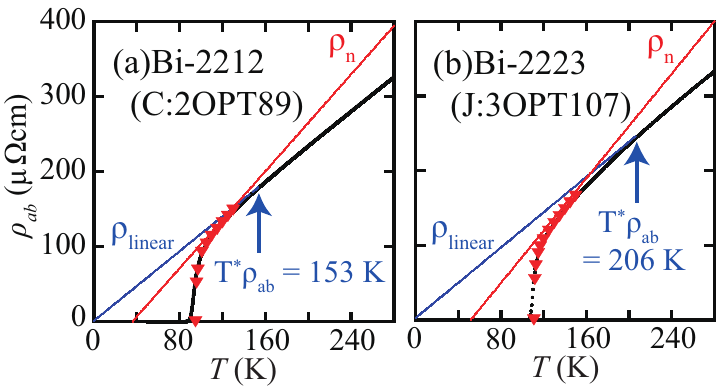}
\caption{\label{f1} (color online). In-plane resistivity, $\rho_{ab}$, for optimally doped (a) Bi-2212 and (b) Bi-2223. Solid straight lines, $\rho_{linear}$, linear extrapolations of $\rho_{ab}$ at higher temperatures, are included to guide the eye. Arrows indicate the temperatures, $T^*_{\rho_{ab}}$, which are defined as the temperatures at which $\rho_{ab}$ decreases by 1\% from the high temperature linear behavior ($\rho_{linear}$). Solid inverted triangles are theoretical fits and $\rho_{n}$ represents the in-plane resistivity when superconductive fluctuation effects are absent.}
\end{center}
\end{figure*}

Figure \ref{f1} (a) and (b) shows the temperature dependence of the zero-field in-plane resistivities, $\rho_{ab}$, for optimally doped Bi-2212 (2OPT89) and Bi-2223 (3OPT107), respectively. In both systems, $\rho_{ab}$ shows a typical downward deviation from high-temperature linear behavior ($\rho_{linear}$) below a certain temperature $T^*_{\rho_{ab}}$. $T^*_{\rho_{ab}}$ is estimated to be 153 and 206 K for 2OPT89 and 3OPT107, respectively. We have attributed this downward deviation to the opening of the pseudogap ~\cite{usu,t.fujii,wata1}. A slightly higher $T^*_{\rho_{ab}}$ in Bi-2223 implies that the inner CuO$_{2}$ plane is less doped compared to the outer CuO$_{2}$ planes~\cite{mukuda}, or that the pseudogap of Bi-2223 is larger than that of Bi-2212~\cite{sato}. Numerical fits using the 2D AL form, $\sigma_{2D-AL}$, for the superconductive fluctuation were performed, dealing with $T_{c0}$ and $\rho_{n}$ as free parameters, and very good fit results were obtained, which are shown in Fig. \ref{f1} (a) and (b). $T_{c0}$ is slightly higher than the observed temperatures at zero resistivity, which may be a reflection of the extreme 2D nature of the sample ~\cite{martin}. One may notice that the slope of $\rho_{n}$ is steeper than that of $\rho_{linear}$ for both Bi-2212 and Bi-2223. Similar analysis has been performed on the samples whose doping levels have been varied and the results are summarized in Table \ref{tb1}. The trend that the slope of $\rho_{n}$ is steeper than that of $\rho_{linear}$ is conspicuous for the underdoped samples. This result reconfirms the fact that $\rho_{ab}$ deviates from high-temperature $T$-linear behavior and decreases rapidly due to the opening of the pseudogap, before being affected by superconductive fluctuation effects upon cooling.

Before conducting the data analysis of the results obtained under magnetic fields, let us briefly show the form of $\sigma^{fl}$. In the IOT theory, $\psi$ is expanded in terms of the Landau orbitals, $\phi_{np_0k_z}$, where $n$ is the quantum number of the Landau levels, $p_0$ denotes the center of the orbitals, and $k_{z}$ is the wavenumber in the magnetic field direction. By computing the current-current correlation function between the Landau orbitals, $\sigma^{fl}$ is obtained as follows~\cite{ikeda} 
\begin{eqnarray}
\sigma^{fl} = \sigma_{0}^{fl} + \delta\sigma^{fl}, \\
\sigma_{0}^{fl} = \frac{e^{2}}{2\hbar\xi_{c}}h^{2}\sum_{n=0}^\infty\frac{n+1}{(\mu_{n+1{R}}-\mu_{n{R}})^{2}}(f_{n}+f_{n+1}-2f_{n+\frac{1}{2}}),
\end{eqnarray}
where
\begin{eqnarray}
f_{n} = \frac{1}{\sqrt{\mu_{n{R}}(1+\frac{d^{2}}{4\xi_{c}^{2}}\mu_{n{R}})}}, \nonumber \\
f_{n+\frac{1}{2}} = \frac{1}{\sqrt{\frac{1}{2}(\mu_{n{R}}+\mu_{n+1{R}})(1+\frac{d^{2}}{4\xi_{c}^{2}}\frac{1}{2}(\mu_{n{R}}+\mu_{n+1{R}}))}}, \nonumber \\
h = (\frac{\xi_{ab}}{r_{0}})^{2}, r_{0} = \sqrt{\frac{\phi_{0}}{2\pi{B}}}. \nonumber
\end{eqnarray}
Based on a tentative estimation, the non Gaussian term, $\delta\sigma^{fl}$~\cite{ikeda}, appearing in Eq. (1) was found to be very small in the temperature region in question. Therefore, this term was neglected throughout our calculations. Here, $\mu_{n{R}}$ is the renormalized ``mass" term of the $n$-th Landau level and is approximately expressed using the renormalized ``mass", $\mu_{0{R}}$, of the lowest Landau levels such that 
\begin{eqnarray}
\mu_{n{R}} \approx \mu_{0{R}}+2nh, \\
\mu_{0{R}} = \mu_{0} + \frac{g_{3}}{\sqrt{\lambda(\beta_{0}^{2}-1)}} +\frac{\lambda\sqrt{\beta_{0}^{2}-1}}{8\beta_{0}}\Biggl[\ln\frac{\gamma_{+}}{\alpha_{+}} \nonumber \\+\frac{\alpha-\beta_{0}}{\sqrt{\beta_{0}^{2}-1}}\ln\Biggl(\frac{\beta_{0}\gamma+\sqrt{(\beta_{0}^{2}-1)(\gamma^{2}-1)}-1}{\beta_{0}\alpha+\sqrt{(\beta_{0}^{2}-1)(\alpha^{2}-1)}-1}\Biggr)\Biggr] , \\
\mu_{0} = \epsilon + h, 
\end{eqnarray}
where,
\begin{eqnarray}
\alpha = 2\beta_{0}^{2}-1, \gamma = \alpha + \frac{8g_{3}\beta_{0}}{\sqrt{\lambda^{3}(\beta_{0}^{2}-1)}}, \nonumber \\
\alpha_{+} = \alpha + \sqrt{\alpha^{2}-1}, \gamma_{+} = \gamma + \sqrt{\gamma^{2}-1}, \nonumber \\
\beta_{0} = 1+\frac{2}{\lambda}\mu_{0{R}}, \lambda = (\frac{2\xi_{c}}{d})^{2}, g_{3} = \frac{k_B}{\Delta{C}}\frac{B}{\phi_{0}\xi_{c}}. \nonumber
\end{eqnarray}

\begin{table*}
\caption{\label{tb1}Several in-plane properties for samples A-E (Bi-2212) and F-K (Bi-2223): $T_c$ (defined by the onset of zero resistivity); pseudogap opening temperature, $T^*\rho_{ab}$; high-temperature $\rho_{ab}$ linear extrapolation, $\rho_{linear}$; normal-state resistivity without superconductive fluctuation, $\rho_{n}$; mean-field transition temperature, $T_{c0}$; and number of holes at Cu atom, $p$. The doping level, $p$, of the Bi-2212 samples was obtained using the empirical relation proposed by Tallon~\cite{tallon}. For Bi-2223, $p$ could not be obtained (see text). Samples are labeled by (i) material, (ii) doping levels, and (iii) $T_c$.}
\begin{center}
\fontsize{7.0pt}{11pt}\selectfont
\begin{tabular*}{150mm}{cccccccccccc}
\hline
Sample & A & B & C & D & E & F & G & H & I & J & K \\
Label & 2UD66 & 2UD70 & 2OPT89 & 2OD79 & 2OD65 & 3UD83 & 3UD90 & 3AS99 & 3AS103 & 3OPT107 & 3OD104
\\ \hline
 $T_c ({K})$ & 65.9 & 69.5 & 89 & 79 & 65 & 83 & 90 & 99 & 103 & 107 & 104 \\
 $T^*\rho_{ab} ({K})$ & 209 & 193 & 153 & - & - & - & 226 & 210 & 206 & 206 & 168 \\
 $\rho_{linear} ({\mu}{\Omega}{cm})$ & 2.28T & 1.6T & 1.16T & - & - & - & 1.5T & 1.75T & 0.7T & 1.18T & 1.18T \\
 $\,$ & +252 & +117 & +2 & - & - & - & +45 & -18 & +20 & +3 & -3 \\
 $\rho_{n} ({\mu}{\Omega}{cm})$ & 3.75T & 2.4T & 1.615T & 1.03T & 1.12T & 2.00T & 2.20T & 2.35T & 1.15T & 1.75T & 1.25T \\
 $\,$ & +35 & +5 & -57 & -27 & +15 & +60 & -90 & -115 & -50 & -90 & -10 \\
 $T_{c0} ({K})$ & 74 & 76.5 & 95 & 83.5 & 68 & 90.5 & 96.2 & 108.2 & 109.5 & 111 & 107.8 \\
 $p ({per Cu})$ & 0.11 & 0.116 & 0.16 & 0.2 & 0.22 & - & - & - & - & - & - \\
\hline
\end{tabular*}
\end{center}
\end{table*}

\begin{figure*}
\begin{center}
\includegraphics{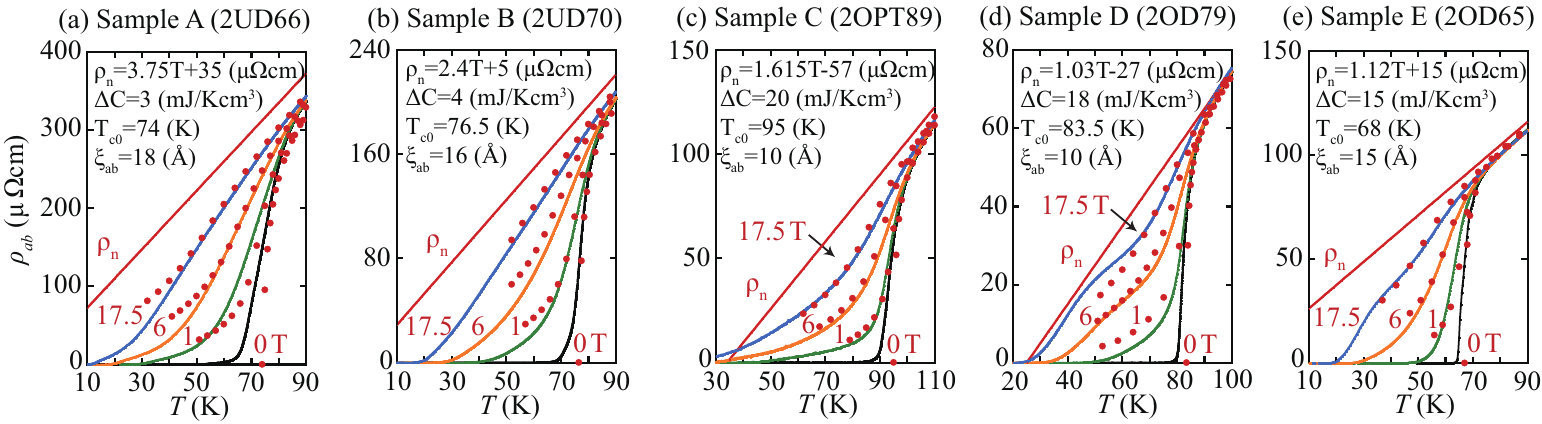}
\caption{\label{f2} (color online). In-plane resistive transitions of single-crystal Bi$_{2}$Sr$_{2}$CaCu$_{2}$O$_{8+\delta}$ (Bi-2212) samples (a) A, (b) B, (c) C, (d) D, and (e) E. Magnetic fields of up to 17.5 T were applied perpendicularly to the CuO$_{2}$ planes. Theoretical fits are shown as solid circles and the parameters used for the fits are shown in the figures.}
\end{center}
\end{figure*}

Figure \ref{f2} (a)-(e) shows the in-plane resistive transition under various magnetic fields for variously doping-controlled Bi-2212. With the fixed $\rho_{n}$ and $T_{c0}$ obtained in the zero-field analysis, theoretical fits using the formula, $\rho_{ab} = 1/(\rho_{n}^{-1} + \sigma^{fl})$, were performed with the adjustable parameters, $\xi_{ab}$ and $\Delta{C}$. The obtained results are also shown in the figure. 
The value of $\Delta{C}$ determines the overall magnitude of the excess conductivity, $\sigma^{fl}$, and therefore determines the gross features at the superconducting transition. On the other hand, the magnitude of $\xi_{ab}$ determines the magnetic-field-dependent fine shape of the transition curve (parallel-shift-like or fan-shaped). The fitting reproduces the characteristic features of the data very well, apart from within the low temperature region, where a vortex motion plays some role in the dissipation. However, this is beyond the range of application for the theory.

\begin{figure*}
\begin{center}
\includegraphics{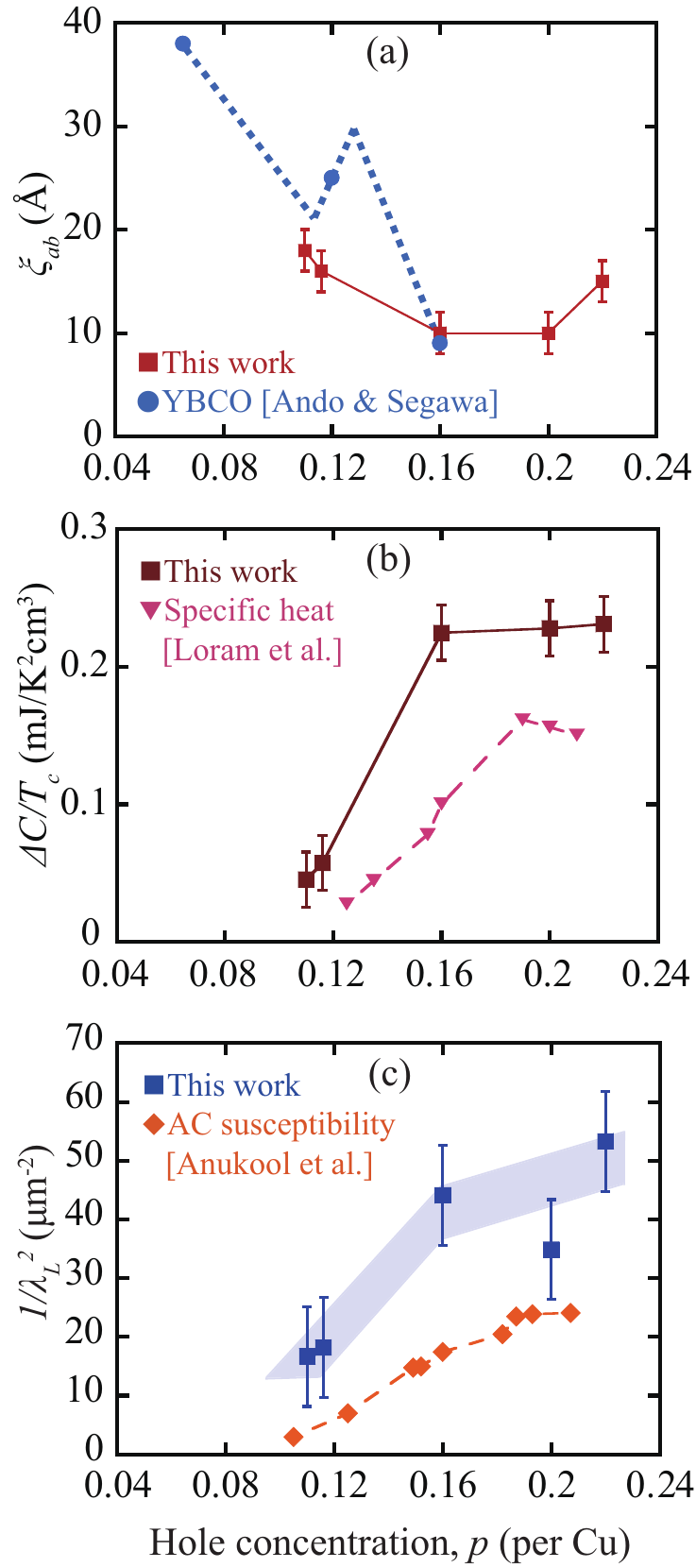}
\caption{\label{f3} (color online).  (a) Plots of $\xi_{ab}$ as functions of $p$ for Bi-2212. The dashed line (solid circles) shows YBCO data~\cite{ando1}.  (b) Plots of $\Delta{C}/T_{c}$ as a function of $p$ for Bi-2212. The dashed line (solid inverted triangles) shows data from specific heat measurement~\cite{loram}. (c) Plots of $1/\lambda_{L}^{2}$ as functions of $p$ for Bi-2212. The dashed line (solid diamonds) shows data from ac susceptibility measurement~\cite{anukool}.}
\end{center}
\end{figure*}

The obtained $\xi_{ab}$ is plotted as a function of $p$ in Figure \ref{f3} (a). Upon doping, $\xi_{ab}$ first decreases in the underdoped region, reaching a minimum in the vicinity of the optimal doping value, and then increases in the overdoped region. Here, we compare the results with those for YBa$_{2}$Cu$_{3}$O$_{y}$ (YBCO)~\cite{ando1} estimated from magnetotransport measurements. At the optimal doping, the $\xi_{ab}$ of both systems are at almost the same value and the doping dependence is quite similar (i.e., $\xi_{ab}$ decreases with increasing doping in the underdoped region) but, in the YBCO case, anomalies around $p = 0.125$ are clearly observed. On the other hand, our results disagree with those generated from Nernst measurements~\cite{wang}. Thus, the doping dependence of $\xi_{ab}$ is not  definitively confirmed at present.

Figure \ref{f3} (b) shows $\Delta{C}/T_{c}$ as a function of $p$. It increases with increased doping up to the optimal value, but it saturates in the overdoped regime. The overall trend of this result agrees with direct thermodynamic measurements~\cite{loram}, apart from the fact that saturation occurs at $p$ = 0.16 in our sample while, according to the thermodynamic data, it occurs at $p$ = 0.19. The reason for this discrepancy in the experimental data is not clear at this stage. By using the obtained values of $\xi_{ab}$, $T_{c0}$, and $\Delta{C}$, $1/\lambda_{L}^{2}$ ($ \propto \rho_{s}$) is estimated via the GL formula shown in the introduction. Figure \ref{f3} (c) shows the results along with those from ac susceptibility measurements~\cite{anukool}. It is found that $1/\lambda_{L}^{2}$ increases with increased doping up to the optimal value and then saturates or slightly increases in the overdoped region, although the data scatters. The doping dependence agrees well with the ac susceptibility measurements, but the obtained values are approximately twice as large as those of ref. 28. The difference might be attributable to the difference in the measurement method or in the sample forms (single crystals vs. aligned powders). The good fitting results produced without the use of a $C$ factor~\cite{oh}, as well as the agreement between the obtained superconducting parameter and the literature values, indicate that the obtained parameters are reasonable. 

In the following, we interpret the obtained results. For the 2D free electrons, the BCS theory gives $\xi_{ab}\propto\hbar{v_{F}}/\Delta_{SG}$. In hole-doped high-$T_{c}$ cuprates, however, the Fermi surface is anisotropic~\cite{shen} and, thus, the $v_{F}$ value should be replaced by the averaged value over the Fermi surface. For simplicity, we assume here that the averaged $v_{F}$ value does not change significantly between different doping states~\cite{zhou}. Consequently, $\xi_{ab}^{-1}$ may be regarded as a measure of $\Delta_{SG}$. In the overdoped region, $\xi_{ab}$ increases upon doping, which is consistent with the observation that $\Delta_{SG}$ decreases~\cite{ding,suzuki1}. In the underdoped region, on the other hand, $\xi_{ab}$ increases with underdoping which, at first glance, contradicts the general observations that $\Delta_{SG}$ increases~\cite{ding,suzuki1}. However, these behaviors may be reconciled as follows. In hole-doped high-$T_{c}$ cuprates (especially in the underdoped states), the pseudogap develops from approximately the ($\pi/a$, 0) and (0, $\pi/a$) direction, resulting in the so-called `Fermi arc'~\cite{norman}. Therefore, the d-wave superconducting gap is restricted to opening mainly on the Fermi arc and the gap amplitude reaches a maximum at the arc edge. Thus, the ``effective" gap, $\Delta_{SG}$ (hereafter, $\Delta_{SG}$ means the ``effective" gap value), is given by $\Delta_{SG} \propto L_{a}\Delta_{0}$~\cite{oda}, where $L_{a}$ is the length of the Fermi arc. Since $L_{a}$ shrinks with decreasing doping~\cite{tanaka}, $\Delta_{SG}$ may decrease, as has been pointed out by Oda et al.~\cite{oda}. This fact explains the increase in $\xi_{ab}$ with decreasing doping.

We note that the analysis performed in this study is based on the superconductive fluctuation theory considering the thermal fluctuation effect only~\cite{ikeda}. For optimally doped or overdoped samples, the dominant fluctuation effect may be the thermal fluctuation effect. However, for heavily underdoped samples near the superconductor-insulator transition, where $\rho_{s}$ is extremely small, it is proposed that a crossover from the thermal to the quantum regime occurs, because of the dominant superconductive fluctuation contribution~\cite{ikeda1}. Indeed, Ikeda has succeeded in explaining the resistivity data obtained under magnetic fields for La$_{2-x}$Sr$_{x}$CuO$_{4}$ (LSCO) over a broad doping range (including the heavily underdoped region) by taking both the quantum superconductive and thermal fluctuation effects into account~\cite{ikeda2}. These reports show that $\xi_{ab}$ decreases with underdoping. Although we believe that the quantum superconductive fluctuation effect is not large for our samples, since the doping levels of our samples are near optimal, the obtained $\xi_{ab}$ results might be modified if the quantum superconductive fluctuation effect were taken into consideration.     

On the other hand, $1/\lambda_{L}^{2}$ is proportional to $\rho_{s}$ and, thus, it is related to the electronic density-of-states (DOS) at the Fermi level. Since pseudogap development depletes the DOS, $1/\lambda_{L}^{2}$ is expected to be reduced with underdoping. This is indeed observed in our experiment and others (Fig. \ref{f3} (c)). In this way, the doping dependences of the superconducting parameters are qualitatively explained by the Fermi arc viewpoint~\cite{yoshida}.  

\begin{figure*}
\begin{center}
\includegraphics[width=150mm]{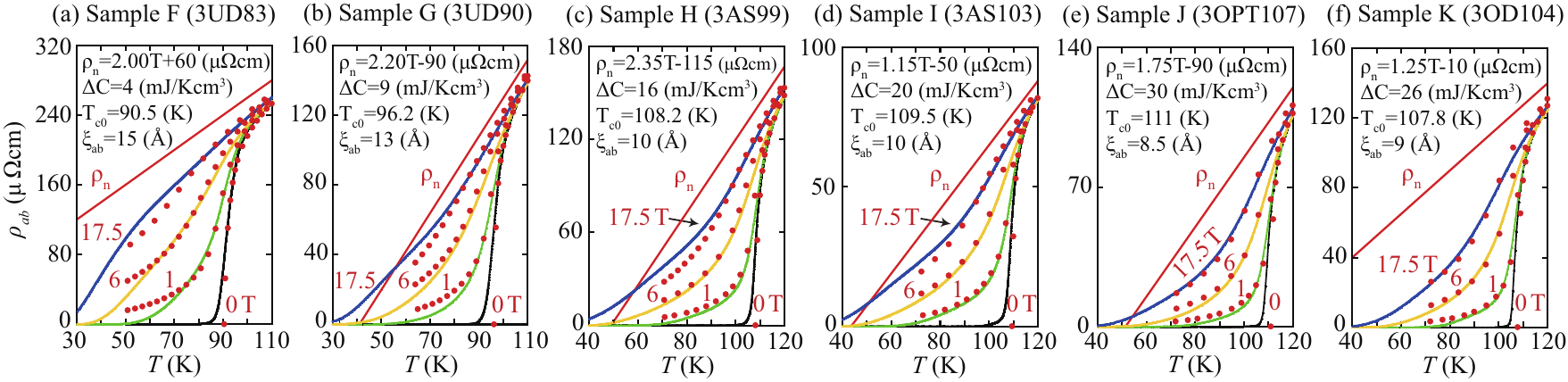}
\caption{\label{f4} (color online). In-plane resistive transitions of single-crystal Bi$_{2}$Sr$_{2}$Ca$_{2}$Cu$_{3}$O$_{10+\delta}$ (Bi-2223) samples (a) F, (b) G, (c) H, (d) I, (e) J, and (f) K. Magnetic fields of up to 17.5 T were applied perpendicularly to the CuO$_{2}$ planes and the theoretical fits are shown as solid circles. The parameters used for the fits are shown in the figures.}
\end{center}
\end{figure*}

Figure \ref{f4} (a)-(f) shows the in-plane resistive transition and the theoretical fits under various magnetic fields for variously doping-controlled Bi-2223. In these figures, the obtained parameters are also shown. It is apparent that the obtained  $\Delta{C}$ monotonically increases with increased doping, up to the optimal point, but it decreases slightly in the overdoped state. On the other hand, $\xi_{ab}$ decreases with increased doping, reaching a minimum at the optimal doping point, and increases slightly in the overdoped state. All these doping-dependent features are qualitatively similar to those of Bi-2212. 

\begin{figure*}
\begin{center}
\includegraphics{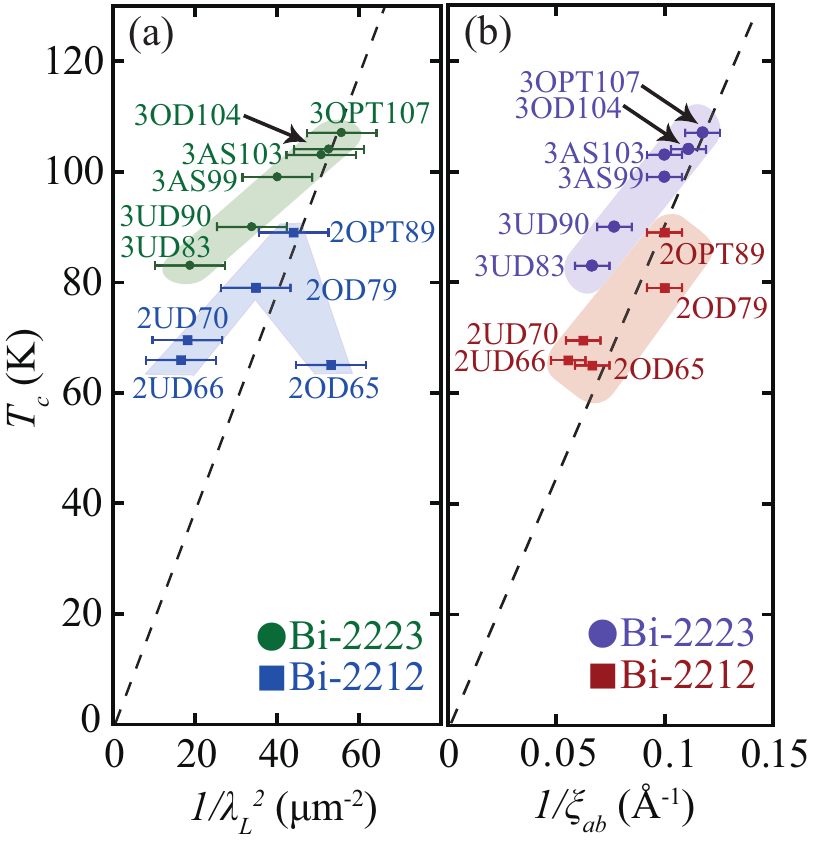}
\caption{\label{f5} (color online). Relationship between superconducting parameters and $T_{c}$ for Bi-2212 (solid squares) and Bi-2223 (solid circles). Plots of (a) $T_{c}$ vs. $1/\lambda_{L}^{2}$ and (b) $T_{c}$ vs. $\xi_{ab}^{-1}$. The dashed lines in each figure represent the scaling relations between the superconducting parameters and $T_{c}$.}
\end{center}
\end{figure*}

To quantitatively compare these parameters with $T_{c}$, plots of $T_{c}$ vs. $1/\lambda_{L}^{2}$ and $T_{c}$ vs. $\xi_{ab}^{-1}$ are shown in Fig. \ref{f5} (a) and (b), respectively, for both Bi-2223 and Bi-2212. We first compare the optimally doped case. The obtained $1/\lambda_{L}^{2}$ values are 55.8 and 44.1 $\mu{m}^{-2}$ for Bi-2223 (3OPT107) and Bi-2212 (2OPT89), respectively. The ratio of these values (55.8/44.1 = 1.27), is close to the $T_{c}$ ratio (107/89 = 1.20). Thus, $1/\lambda_{L}^{2}$ ($\propto$ $\rho_{s}$) scales with $T_{c}$, which is consistent with the results of the ARPES measurements~\cite{feng1}. The obtained $\xi_{ab}$ values are 8.5 and 10 {\AA}  for 3OPT107 and 2OPT89, respectively and the inverse ratio (10/8.5 = 1.18) is therefore close to the $T_{c}$ ratio. Thus, $\xi_{ab}^{-1}$ also scales with $T_{c}$. If the relation of $\xi_{ab}^{-1} \propto $ $\Delta_{SG}$ holds, this result is consistent with the ARPES results~\cite{sato,matsui,ideta,feng1}.

We next compare the other doping states. One can clearly see that $T_{c}$ vs. $1/\lambda_{L}^{2}$ for both compounds do not fall in one line (Fig. \ref{f5} (a)). Moreover, in the underdoped region, the data for each compound does not extrapolate to the origin. These results imply that the ``Uemura's plot" ($T_{c} \propto \rho_{s}$) condition is not generally satisfied. This fact has already been pointed out by Tallon et al.~\cite{tallon1}. That is, $T_{c}$ is not determined by $\rho_{s}$ only, and the related phase fluctuation model~\cite{emery} may not explain the behavior of $T_{c}$, even in the underdoped region. On the other hand, as can be seen in Fig. \ref{f5} (b), the Bi-2212 data on $\xi_{ab}^{-1}$ roughly satisfies the scaling relation. However, the Bi-2223 data deviates from this simple scaling behavior and is always above that of Bi-2212. Therefore, the pairing strength represented by $\xi_{ab}^{-1}$ does not solely determine $T_{c}$.

In order to further understand this result, let us compare sample G (3UD90) and sample C (2OPT89). Sample G has much smaller $1/\lambda_{L}^{2}$ ($\rho_{s}$) (Fig. \ref{f5} (a)) and longer $\xi_{ab}$ (thus, smaller $\Delta_{SG}$) (Fig. \ref{f5} (b)) compared with those of sample C, implying sample G is at an obvious disadvantage to Sample C as regards the realization of high-$T_c$ superconductivity. However, the $T_{c}$ values of both samples are almost the same. This fact suggests that sample G acquires additional superconducting condensation energy compared to sample C, from some source other than the in-plane $\Delta_{SG}$ (or $\rho_{s}$). The mechanism of this process is an open question, however, one possible source is the interlayer tunneling mechanism~\cite{chak,anderson}. In this model, the kinetic energy in the normal state is increased, either due to a non-Fermi-liquid nature or opening of the pseudogap, while in the superconducting state, the kinetic energy is restored by the interlayer Josephson coupling, resulting in enhanced Cooper pairing. Thus, the increase in the number of CuO$_{2}$ planes results in an increase in the superconducting condensation energy. 
To verify this mechanism, more comprehensive studies using various experimental techniques are needed in the future.

\section{Conclusions}

In summary, magnetotransport measurements on Bi-2212 and Bi-2223 have been systematically performed, and superconducting parameters such as $\xi_{ab}$ and ${\Delta}C$ have been reasonably extracted from theoretical fits of the transition curves. The doping dependence of these superconducting parameters has been qualitatively explained using the Fermi arc approach. However, neither the pairing strength (represented by $\xi_{ab}^{-1}$) nor the phase stiffness ($\rho_{s}$) explains the behavior of $T_c$. Instead, the $n$-dependence of these parameters suggests that an additional superconducting condensation energy exists, due to the increase in the number, $n$, of CuO$_{2}$ planes. A successful model for high-$T_c$ superconductivity should consistently account for these experimental observations, as well as the NMR results (the coexistence of the antiferromagnetic order and the high-$T_c$ superconductivity).

\begin{acknowledgment}
We thank R. Ikeda, H. J. Im, J. Goryo, H. Nakazawa, S. Awaji, and K. Semba for helpful input. The magnetoresistance measurements were performed at the High Field Laboratory for Superconducting Materials, Institute for Materials Research, Tohoku University. This work was supported by JSPS KAKENHI Grant Number 25400349.
\end{acknowledgment}


\end{document}